\documentclass[11pt,a4paper]{article}
\usepackage{jheppub}

\usepackage{amsfonts}
\usepackage{slashed}

\def\bea{\begin{eqnarray}}
\def\eea{\end{eqnarray}}

\def\lmatrix{\left(\begin{array}}
\def\rmatrix{\end{array}\right)}

\def\msbar{\overline{\rm MS\kern-0.5pt}\kern0.5pt}

\title{The flavor dependence of $m_\varrho / f_\pi$}

\author{Daniel Nogradi}
\author{and Lorinc Szikszai}

\affiliation{E\"otv\"os University, Institute for Theoretical Physics, Budapest 1117, Hungary}

\emailAdd{nogradi@bodri.elte.hu}
\emailAdd{szikszail@caesar.elte.hu}

\abstract{We calculate the $m_\varrho / f_\pi$ ratio in the chiral and continuum limit for $SU(3)$ 
gauge theory coupled to $N_f = 2,3,4,5,6$ fermions in the fundamental representation. 
Keeping all systematic effects under full control we find
no statistically significant $N_f$-dependence; $m_\varrho / f_\pi = 7.95(15)$.
Assuming the KSRF-relations we conclude that 3 other low energy
quantities related to the vector meson are also $N_f$-independent within errors including the $\varrho\pi\pi$ coupling
$g_{\varrho\pi\pi}$. If the model is thought of as a strong
dynamics inspired composite Higgs model our results indicate that the experimentally most easily accessible 
new composite particle, the
vector meson, and its properties may be robust and independent of the fermion content of the model as long as the gauge
group is $SU(3)$, provided $N_f$-independence extends all the way to the conformal window.}


\arxivnumber{}

\begin{document}

\maketitle

\section{Introduction and summary}
\label{intro}

The possibility of a composite Higgs boson disguised as a scalar resonance in a so far unobserved 
strongly interacting gauge sector led to renewed interest in lattice calculations in models with unusual fermion content. 
As the fermion content varies for a given gauge group the non-perturbative dynamics of gauge theory changes drastically.
If the fermion representation is also fixed the fermion content is controlled by the flavor number $N_f$. As $N_f$
increases but stays below the conformal window 
the number of Goldstone bosons increases, the $\beta$-function decreases in magnitude and hence the running
becomes slower,
the topological susceptibility
decreases at fixed Goldstone mass and decay constant, etc. Change in the infrared dynamics as $N_f$ is approaching the
conformal window is expected since an even more drastic change will eventually occur as $N_f$ passes into the conformal
window. Yet there are hints from past lattice calculations of $SU(3)$ gauge theory 
that one particular ratio $m_\varrho / f_\pi$ in the chiral limit is surprisingly
stable as $N_f$ varies. The available results are at finite lattice spacing which makes their comparison hard and 
finite volume effects are not always negligible but there are
indications that $m_\varrho / f_\pi \sim 8.0$ for $N_f = 2, 4, 6, 8, 9$ with fundamental fermions 
\cite{Fodor:2009wk, Jin:2009mc, Aoki:2013xza, Jin:2013hpa, Fleming:2013tra, Appelquist:2016viq, Appelquist:2018yqe}
and even with $N_f = 2$
sextet fermions \cite{Fodor:2012ty, Fodor:2016pls, Fodor:2016wal}. Not to mention the value for QCD $\sim 8.4$ 
which is also not far even though the quarks are massive.

In this work we aim to study the ratio $m_\varrho / f_\pi$ more systematically. 
Our goal is to obtain controlled continuum results for $m_\varrho / f_\pi$ in the chiral limit with $SU(3)$ and $N_f =
2,3,4,5,6$ in order to see the continuum $N_f$-dependence, if any. First, for each $N_f$ we have carefully determined 
the size of finite volume effects and quantified how large $m_\pi L$ needs to be in order to have only sub-percent
distortions from the finite volume. As expected $m_\pi L$ needs to grow with $N_f$, more specifically a linear
relationship is found, $m_\pi L$ needs to increase linearly in order to maintain at most 1\% finite volume effect. 
For each model, i.e. fixed $N_f$ we then simulate at 4 lattice spacings and
4 fermion masses at each always ensuring that finite volume effects are below 1\%. 
The 16 simulation points per $N_f$ allow for fully controlled chiral-continuum extrapolations 
leading to our final results in figure \ref{mrhofpinf} which indeed shows no statistically significant
$N_f$-dependence, a constant fit as a function of $N_f$ leads to $m_\varrho / f_\pi = 7.95(15)$.

This remarkable $N_f$-independence is not at all trivial and is not guaranteed by any general principle as far as we are
aware. It should be noted that the celebrated KSRF-relations \cite{Kawarabayashi:1966kd, Riazuddin:1966sw} 
do state non-trivial relationships among 
various $\varrho$-related low energy quantities based on phenomenological assumptions but they do not say anything about
their $N_f$-dependence. In theory the KSRF-relations (see section \ref{conclusionandoutlook} for details) 
may hold to high precision at each $N_f$ and the quantities
themselves may very well vary with $N_f$. The fact that this does not happen seems to be a non-trivial property of
$SU(3)$ gauge theory. On the other hand assuming the KSRF-relations our results lead to $N_f$-independence of the
$\varrho\;\pi\;\pi$ coupling $g_{\varrho\pi\pi}$,  $\Gamma_\varrho /  m_\varrho$ where $\Gamma_\varrho$ is the width,
and $f_\varrho / m_\varrho$ where $f_\varrho$ is the decay constant.

Our original motivation was the study of composite Higgs models with gauge group $SU(3)$. In the class of models we have
in mind the Higgs boson is identified as the $O^{++}$ scalar flavor singlet meson and the scale is set by 
$f_\pi = 246\;GeV$. Our results then mean that the mass of the vector
resonance which is the experimentally most easily accessible new particle prediction is at $\sim 2\;TeV$ regardless of
what the fermion content is.

Beside the beyond Standard Model motivation we believe the ratio $m_\varrho / f_\pi$ will be useful in understanding the
dynamics of crossing into the conformal window. Inside the conformal window all masses are vanishing of course. It is
possible however to define the conformal models at finite fermion mass and then $m_\varrho$, $f_\pi$ and all other finite
renormalization group invariant dimensionful quantities will scale to zero with a common power of the fermion
mass, leading to a well-defined ratio $m_\varrho / f_\pi$ in the massless limit even inside the conformal window. For
example in the free theory, corresponding to $N_f = 33/2$, we have $m_\varrho = 2m$ and $f_\pi = \sqrt{12} m$ where $m$
is the fermion mass \cite{Cichy:2008gk} and obtain $m_\varrho / f_\pi = 1 / \sqrt{3}$. 
It may be the case that $m_\varrho / f_\pi$ stays
flat all the way to the conformal window as $N_f$ grows and then gradually drops to $1/\sqrt{3}$ at $N_f = 33/2$. Or it
may be that a more abrupt change occurs at the lower end of the conformal window. We leave these speculations to future
work.

The organization of the paper is as follows. In section \ref{simulationdetails} our choice of discretization is
described and the details of the simulation, in section \ref{finitevolumeeffects} the detailed study of finite volume
effects as a function of $N_f$ is given. Section \ref{chiralcontinuumextrapolation} contains the main results of our
work, the chiral and continuum extrapolation of the ratio $m_\varrho / f_\pi$ for all $N_f$ as well as the topological
susceptibility. The latter is used to test for the appropriate $O(a^2)$ scaling of taste breaking effects inherent to
staggered fermions. Finally section \ref{conclusionandoutlook} ends with our conclusions and future outlook.

\section{Simulation details}
\label{simulationdetails}

The numerical simulations use the Symanzik tree-level improved gauge action and 4 steps of stout improved
\cite{Morningstar:2003gk, Durr:2010aw} staggered
fermions with smearing parameter $\varrho = 0.12$.
This particular choice of action has been shown to have relatively small cut-off effects in both small and large
physical volume simulations \cite{Fodor:2012td,Fodor:2015baa, Fodor:2015zna, Fodor:2017die}.
The $N_f = 4$ case requires no rooting of the staggered determinant and the HMC
\cite{Duane:1987de} algorithm
is used. The other flavor numbers use either RHMC \cite{Clark:2006fx} 
only ($N_f = 2, 3$) or a combination of HMC and RHMC ($N_f = 5, 6$) in
order to have the correct number of continuum flavors. Multiple time scales \cite{Sexton:1992nu} and Omelyan integrator 
\cite{Takaishi:2005tz} are used to speed up the simulations. On all lattices the temporal extent is twice the spatial
extent $L/a$.

\begin{figure}
\begin{center}
\includegraphics[width=7.5cm]{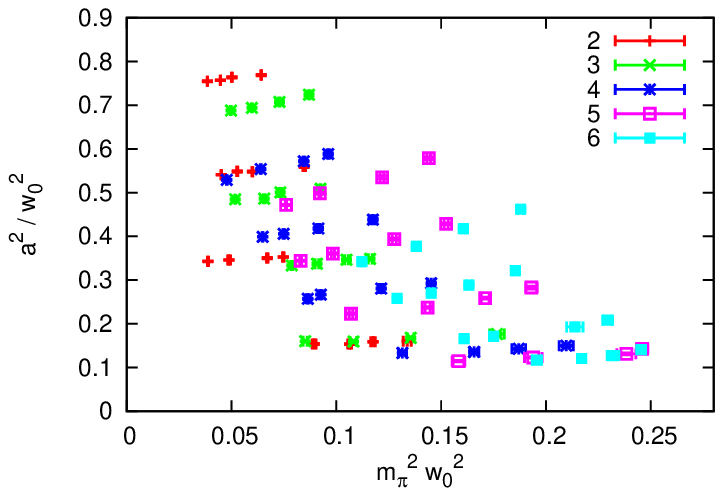} \includegraphics[width=7.5cm]{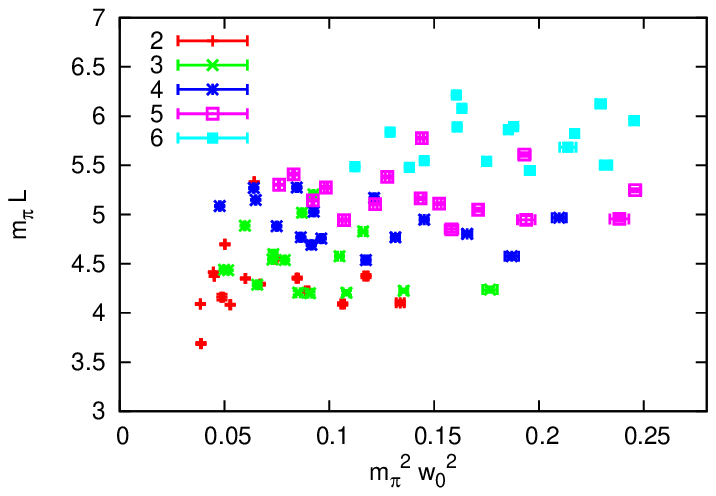} 
\end{center}
\caption{The landscape of simulation points for the various flavor numbers $N_f$. Left: cut-off and Goldstone mass. Right: volume and Goldstone
mass.}
\label{landscape}
\end{figure}

The observables we measure are $m_\pi, f_\pi, m_\varrho, w_0$ and the topological susceptibility $Q^2/V$. The
scale $w_0$ \cite{Borsanyi:2012zs} is measured using the SSC discretization according to the terminology in \cite{Fodor:2014cpa}.
For each $N_f$ simulations were carried out at four lattice spacings
and four fermion masses at each lattice spacing giving a total of 16 points; these are tabulated in tables \ref{data234} and
\ref{data56}. The total number of thermalized
trajectories is $1000-2500$ and every $10^{th}$ is used for measurements.

The landscape of simulation points in terms of cut-off, Goldstone mass and volume is shown in figure \ref{landscape}.

\section{Finite volume effects}
\label{finitevolumeeffects}

The simulations are performed in finite volume and the associated systematic errors ought to be controlled.
In order to
have fully controlled finite volume effects two issues need to be addressed. One, it is important to be in 
the kinematical regime where the $\varrho$-meson can
not decay into pions. Hence all simulations were performed in the regime such that $m_\varrho/(2m_\pi) < \sqrt{1 +
\left(\frac{2\pi}{m_\pi L}\right)^2 }$. This constraint mainly prevents us from reaching too light fermion masses at
rough lattice spacings. 
Two, the topological charge should fluctuate enough and should not be frozen so as not to have
approximately fixed topology simulations. This constraint is most relevant at small lattice spacings and we made
sure that topology does change frequently enough even at the finest lattice spacings we use. The topology change is
frequent enough such that we are able to measure the topological susceptibility for all runs and the expectations from
tree level chiral perturbation theory are confirmed (see next section).

\begin{figure}
\begin{center}
\includegraphics[width=5.4cm]{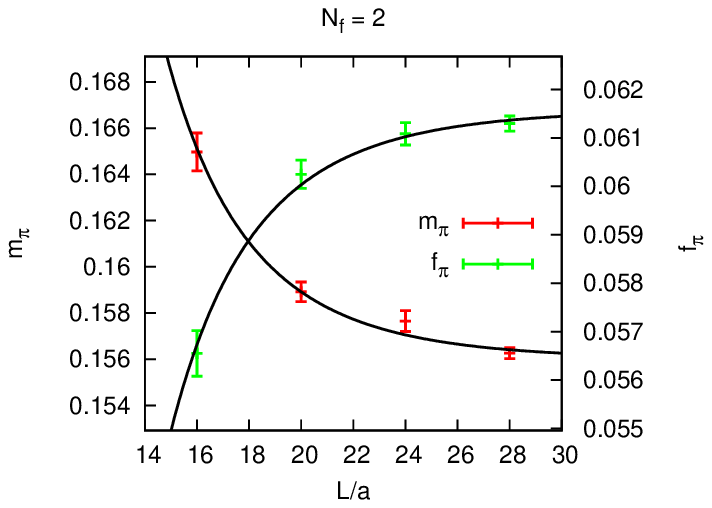} \hspace{1cm} \includegraphics[width=5.4cm]{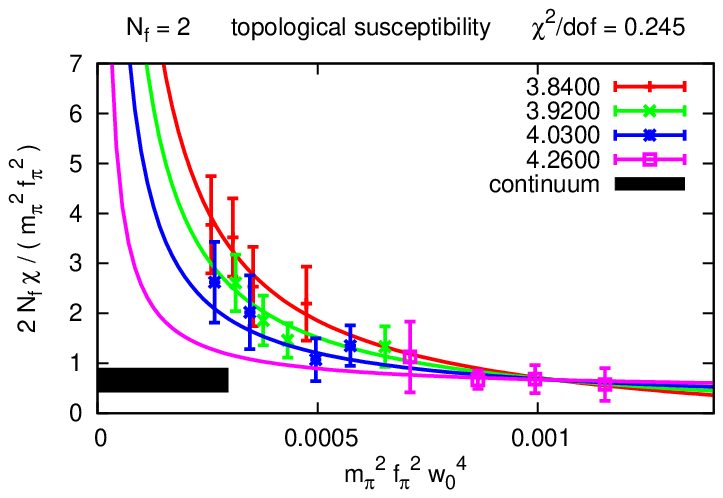} \\
\includegraphics[width=5.4cm]{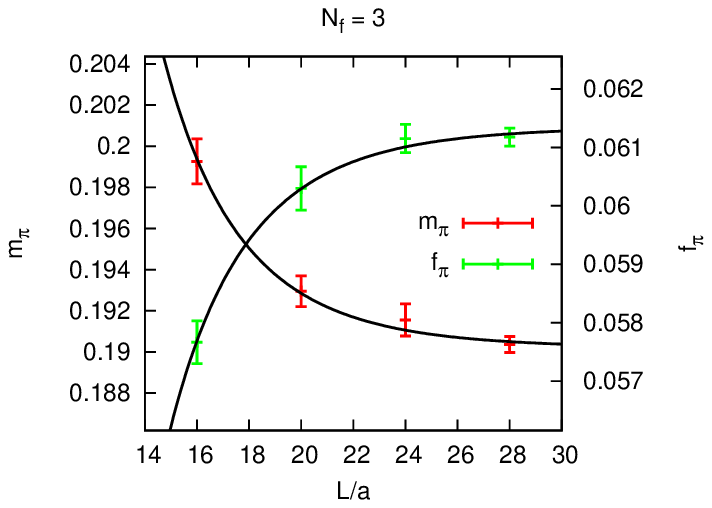} \hspace{1cm} \includegraphics[width=5.4cm]{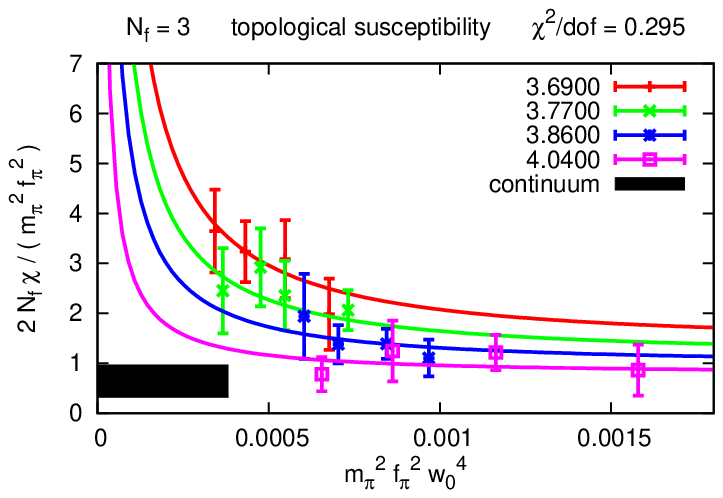} \\
\includegraphics[width=5.4cm]{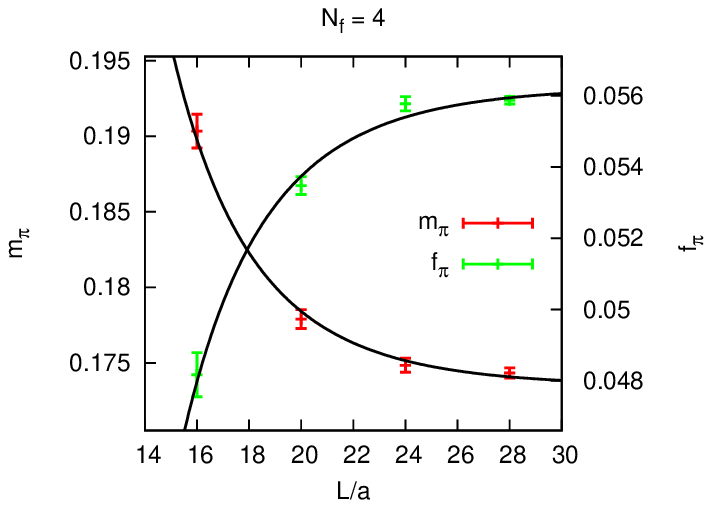} \hspace{1cm} \includegraphics[width=5.4cm]{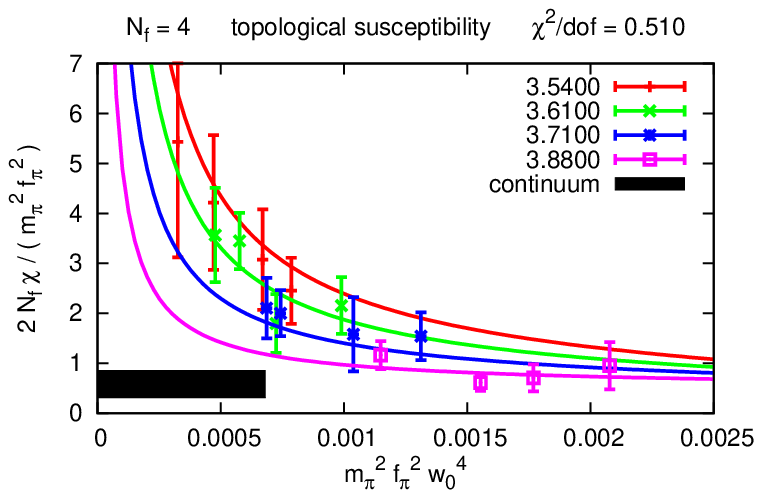} \\
\includegraphics[width=5.4cm]{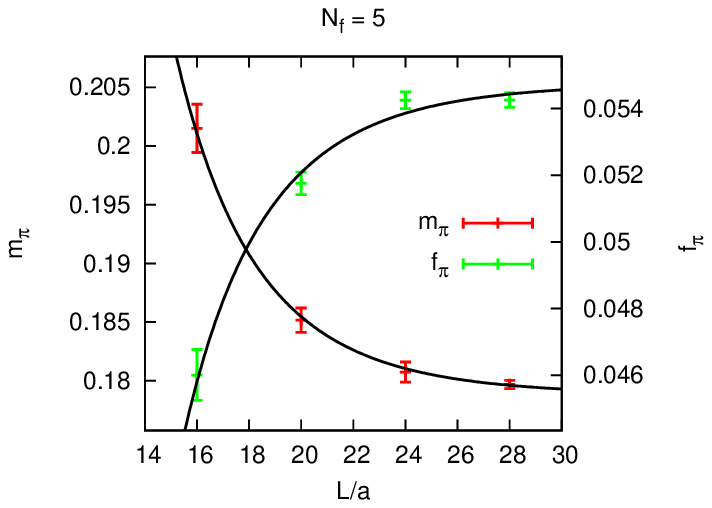} \hspace{1cm} \includegraphics[width=5.4cm]{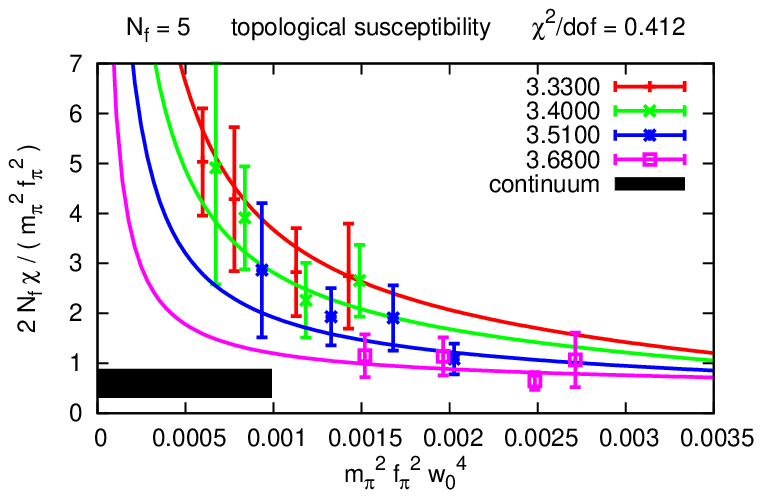} \\
\includegraphics[width=5.4cm]{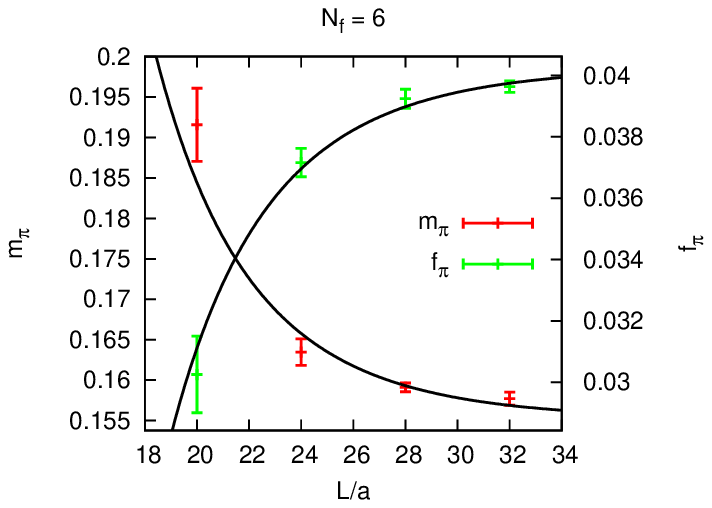} \hspace{1cm} \includegraphics[width=5.4cm]{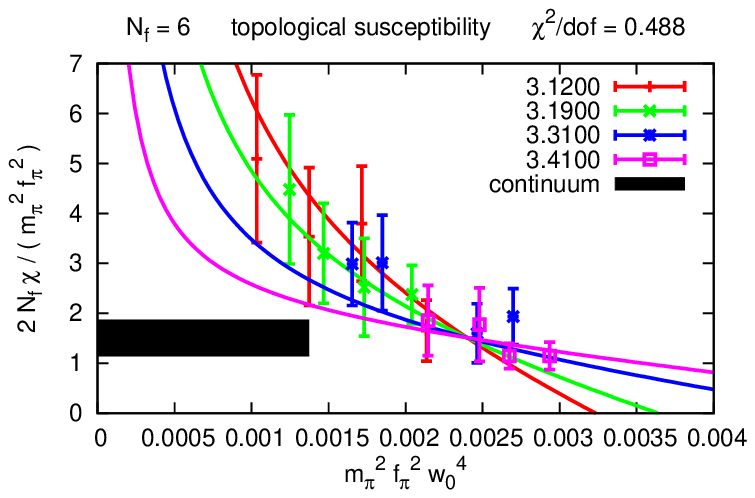} \\
\end{center}
\caption{Left: infinite volume extrapolations of $m_\pi$ and $f_\pi$, based on table \ref{finvoltable}.
Right: chiral-continuum extrapolation of the topological susceptibility. The ratio $2 N_f \chi / ( m_\pi^2 f_\pi^2)$
is shown which is expected to be constant $1$ at leading order of chiral perturbation theory;
see text for more details.}
\label{finvolplottopsusc}
\end{figure}

Once these two issues are handled properly the finite volume effects in $m_\pi$ and $f_\pi$ 
are purely exponential in $m_\pi L$. The residual finite volume effect in $m_\varrho$ can be estimated based on the relationship
between finite volume energy levels and scattering states \cite{Luscher:1990ux, Luscher:1991cf}. 

How large $m_\pi L$ needs to be in order to have a fixed small finite volume effect in $m_\pi$ and $f_\pi$, 
e.g. less than $1\%$, depends on
$N_f$. For each flavor number we have performed dedicated finite volume runs at fixed lattice spacing and fermion mass.
The infinite volume extrapolation is through the non-linear fit
\bea
m_\pi(L) &=& m_{\pi\infty} + C_m\; g(m_{\pi\infty}L)
\eea
where the fit parameters are $m_{\pi\infty}$ and $C_m$. The form of the finite volume correction \cite{Gasser:1986vb}  
is given by
\bea
g(x) &=& \frac{4}{x} \sum_{n\neq 0} \frac{K_1(nx)}{n}
\eea
with the modified Bessel function of the second kind $K_1$ and the sum is over integers $(n_1,n_2,n_3,n_4)$ such that 
$n^2 = n_1^2 + n_2^2 + n_3^2 + 4n_4^2 \neq 0$; see also \cite{Colangelo:2005gd}.
The sum may be replaced by the first exponential and all infinite volume
extrapolations were repeated as a cross-check with a single exponential and give identical results, within errors, to
the one obtained using the full $g(x)$ function.

Using the $m_\pi(L)$ data the infinite volume extrapolated $m_{\pi\infty}$ and its error may be obtained. Once this is
done the decay constant $f_\pi(L)$ needs to be extrapolated as well, using a similar expression \cite{Gasser:1986vb},
\bea
f_\pi(L) &=& f_{\pi\infty} - C_f \; g(m_{\pi\infty}L)
\eea
where now the fit is linear in the fit parameters $f_{\pi\infty}$ and $C_f$. The statistical error on $m_{\pi\infty}$
does need to be propagated carefully into the above fit of course. Note that $C_f > 0$ and $C_m > 0$,
i.e. masses decrease towards larger volumes while the decay constant increases. The net effect on the ratio $m_\varrho /
f_\pi$ is an enhancement of finite volume effects.

Our results for all flavor numbers are shown in figure \ref{finvolplottopsusc} (left) 
based on the data in table \ref{finvoltable}.
The main conclusion is that in order to have a fixed small finite volume effect, $m_\pi L$ needs to grow linearly with
$N_f$. For instance in order to have less than $1\%$ finite volume effects in $m_\pi$ and $f_\pi$ the following needs to
hold for the spatial volume,
\bea
m_\pi L > 3.10 + 0.35 N_f \;.
\label{mpil}
\eea
Clearly, as $N_f$ is increasing finite volume effects get larger. The conventional rule of thumb $m_\pi L > 4$ from QCD
is satisfactory for sub-percent finite volume effects at $N_f = 2, 3$ but for $N_f = 4,5,6$ it is not. For instance at
$N_f = 6$ one needs $m_\pi L > 5.2$.
As the fermion content gets larger and the model moves closer to the conformal
window finite volume effects grow, in line with general expectations. Here we have quantified this
phenomenon for $SU(3)$ and fundamental fermions.

In our runs condition (\ref{mpil}) is satisfied which means that our chiral extrapolations are essentially in
infinite volume. The chiral expansion in infinite volume \cite{Gasser:1984gg} is indeed applicable to the 
simulations once $f_\pi L$ is large enough. For all our simulations we have $ 0.95
\leq f_\pi L \leq 1.92$. Note that our $f_\pi$ is in the ``lower'' convention, i.e. the one which gives 
$f_\pi \approx 92\;MeV$ and not $130\;MeV$, in QCD.

Finite volume effects in $m_\varrho$ can be estimated a posteriori as follows.
Using the KSRF-relation (\ref{ksrfrelation}) we determine $g_{\varrho\pi\pi}$ for each $N_f$.
Once $g_{\varrho\pi\pi}$ is fixed the finite volume effects
are given by the relationship between finite volume energy levels and scattering states \cite{Luscher:1990ux,
Luscher:1991cf} and $m_\varrho$ in infinite volume can be obtained from a single volume as is the case for us. The
result here is that finite volume effects are at most 2\% at 12 out of the 16 simulation points at $N_f = 2$,
are at most 2\% at all 16 simulation points at $N_f = 3$, are at most 1\% at 15 out of the 16 simulation points at $N_f
= 4$ and at most 1\% for all points at $N_f = 5$ and $N_f = 6$. Our final statistical uncertainties for $w_0 m_\varrho$
are between 3\% and 5\% hence we conclude a posteriori that finite volume effects for all of our observables are under control.

\section{Chiral-continuum extrapolation}
\label{chiralcontinuumextrapolation}

Before discussing the chiral-continuum extrapolations it is worth remembering that staggered fermions, as is well-known, 
suffer from taste breaking. This means that the measured pseudo-scalar meson is the lightest of the full taste broken
multiplet and the higher ones ($N_f^2-2$ of them) 
do not chirally extrapolate to zero at fixed non-zero lattice spacing. In other words the
chiral $SU(N_f) \times SU(N_f)$ group is broken at finite lattice spacing and the $N_f$-dependence of low energy
observables on the lattice is not necessarily the same as in the continuum, only if the lattice spacing is small
enough and all Goldstone bosons are light enough. Hence before attempting to extrapolate both chirally and to the continuum
our main observable, the $m_\varrho / f_\pi$ ratio, we sought a quantity which is as sensitive to $N_f$ as possible
in order to test whether our simulations are close enough to the continuum and zero fermion mass limit.

\subsection{Topological susceptibility}

A powerful test of whether at finite lattice spacing the effective number of light degrees of freedom is the
same as in the continuum is given by the topological susceptibility. 
The topological susceptibility is very sensitive to
the light degrees of freedom since these are the ones at small fermion mass which suppress non-zero topology.
As a result $N_f$-dependence shows up already at the leading order of chiral perturbation theory
\cite{Leutwyler:1992yt},
\bea
\label{chi}
\frac{\langle Q^2 \rangle}{V} = \chi = \frac{1}{2N_f} f_\pi^2 m_\pi^2
\eea
i.e. the $N_f$-dependence is fixed once $m_\pi$ and $f_\pi$ are measured. 
We have performed a combined chiral-continuum extrapolation of the topological susceptibility
for each $N_f$ and have confirmed the above expectation, indicating
that the light degrees of freedom are correctly captured, i.e. any deviation from the continuum due to
taste broken Goldstone bosons is correctly
extrapolated to zero as $O(a^2)$. This is a highly non-trivial test for each $N_f$ and we take it to indicate that the
lattice spacings and bare fermion masses were indeed chosen such that a combined chiral-continuum extrapolation is
meaningful.

More precisely, we use the gradient flow based \cite{Luscher:2010iy} discretization of the topological charge, measuring it
at $t=w_0^2$. Note that the chiral extrapolation of $\chi$ at finite lattice spacing does not need to vanish,
precisely because of the fact that the taste broken Goldstone bosons do not extrapolate to zero \cite{Billeter:2004wx}. 
Hence we adopt the following combined chiral-continuum extrapolation at each $N_f$,
\bea
\label{chifit}
\chi w_0^4 = C_0  m_\pi^2 f_\pi^2 w_0^4  + C_1 \frac{a^2}{w_0^2} + C_2 \frac{a^2}{w_0^2} ( m_\pi^2 f_\pi^2 w_0^4 )\;,
\eea
where the fit parameters are $C_0, C_1$ and $C_2$.
The continuum expectation (\ref{chi}) is then $C_0 = 1/(2N_f)$. In figure \ref{finvolplottopsusc} (right) we plot the ratio
$2 N_f \chi / ( m_\pi^2 f_\pi^2 )$ for each $N_f$ which ought to be consistent with the constant $1$ in the chiral
continuum limit. 
At each $\beta$ we also fit $a^2/w_0^2$ as a linear function of $m_\pi^2f_\pi^2 w_0^4$ and then using the fitted
$C_{0,1,2}$ coefficients together with (\ref{chifit}) 
we also show the resulting mass dependence at each $\beta$ by the solid lines in order to get a
sense of the size of cut-off effects.
The extrapolated $2 N_f C_0$ coefficient is shown by the black bands, these are $0.67(24)$, $0.65(33)$,
$0.59(28)$, $0.60(29)$ and $1.51(36)$ for $N_f = 2,3,4,5,6$, respectively.
We find that for all $N_f$ there is agreement with $1$ within at most $1.5\,\sigma$, which is even though not perfect, 
certainly better than expected since we only fit the leading order expression.
All $\chi^2/dof$ (with $dof=13$) of the extrapolations are below unity.

\subsection{The $m_\varrho / f_\pi$ ratio}

Having quantitative confirmation that taste breaking effects scale to zero as $O(a^2)$ as expected we turn to the main
object of our study, the $m_\varrho / f_\pi$ ratio. In order to estimate the systematic error coming from the
chiral-continuum extrapolation we have performed two types of fits.

\begin{figure}
\begin{center}
\includegraphics[width=5.1555cm]{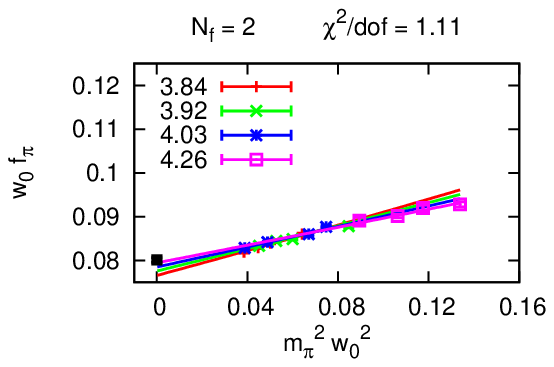} \hspace{-0.4cm} \includegraphics[width=5.1555cm]{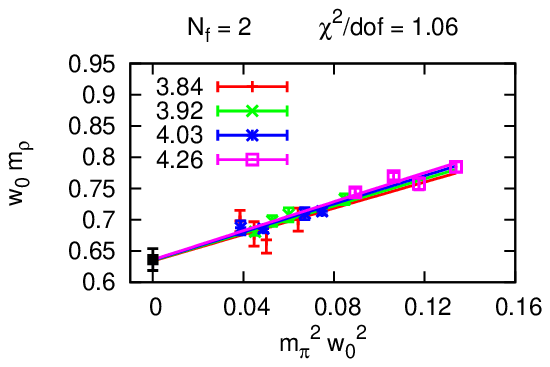} \hspace{-0.4cm} \includegraphics[width=5.1555cm]{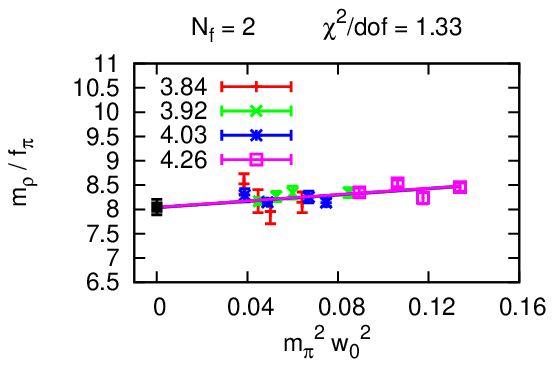} \\
\includegraphics[width=5.1555cm]{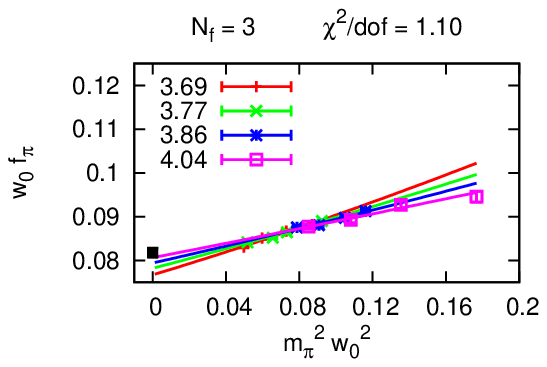} \hspace{-0.4cm} \includegraphics[width=5.1555cm]{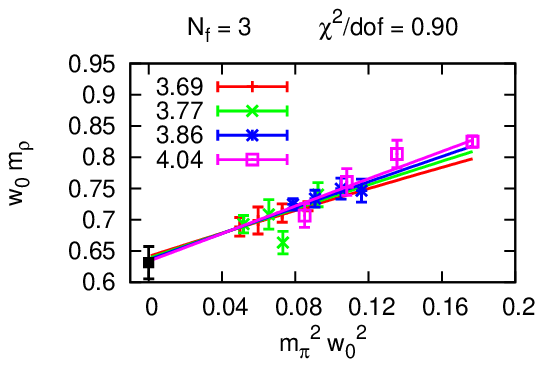} \hspace{-0.4cm} \includegraphics[width=5.1555cm]{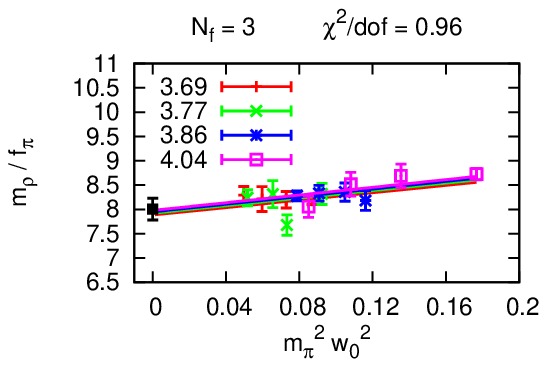} \\
\includegraphics[width=5.1555cm]{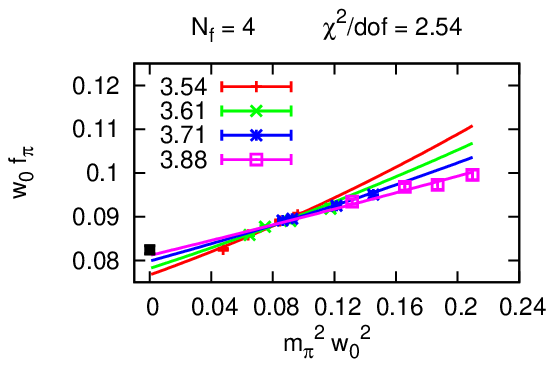} \hspace{-0.4cm} \includegraphics[width=5.1555cm]{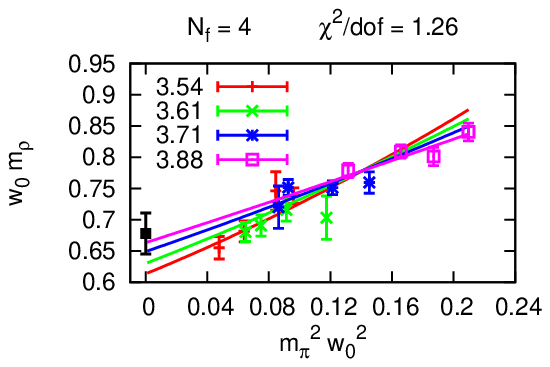} \hspace{-0.4cm} \includegraphics[width=5.1555cm]{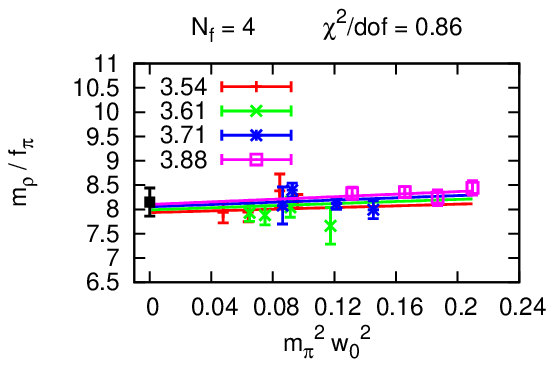} \\
\includegraphics[width=5.1555cm]{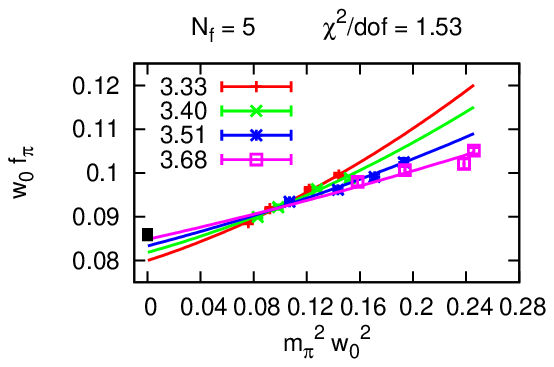} \hspace{-0.4cm} \includegraphics[width=5.1555cm]{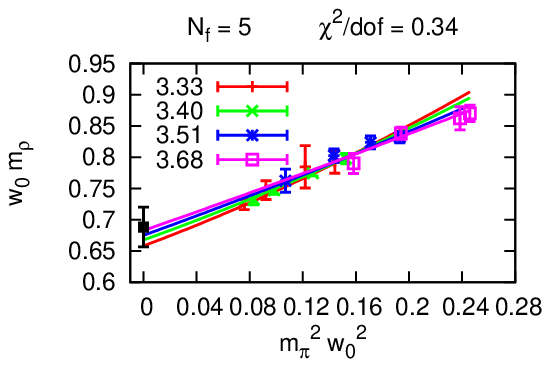} \hspace{-0.4cm} \includegraphics[width=5.1555cm]{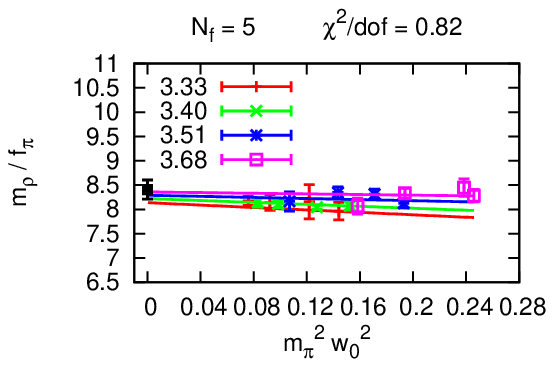} \\
\includegraphics[width=5.1555cm]{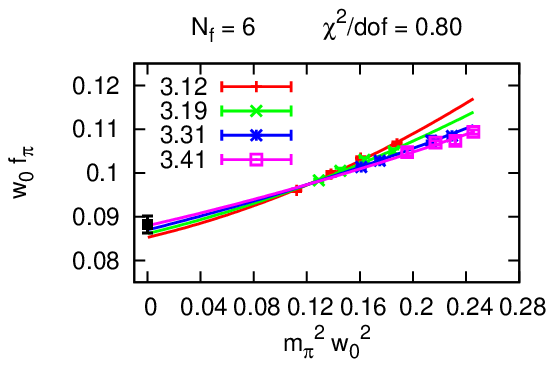} \hspace{-0.4cm} \includegraphics[width=5.1555cm]{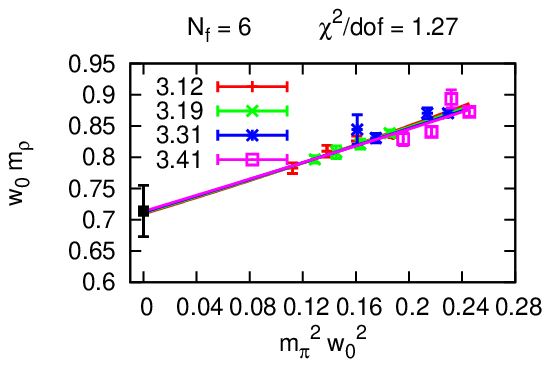} \hspace{-0.4cm} \includegraphics[width=5.1555cm]{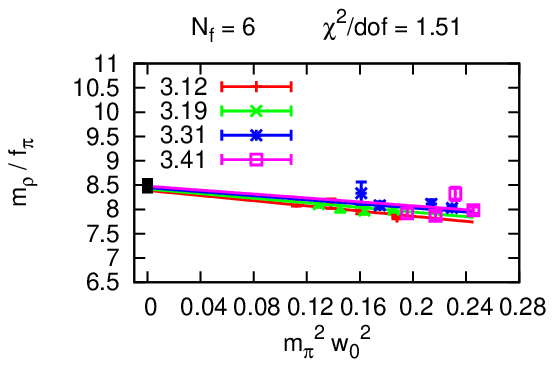} 
\end{center}
\caption{Chiral-continuum extrapolation of $f_\pi w_0$,  $m_\varrho w_0$ (left two columns) 
and directly the ratio $m_\varrho / f_\pi$ (right most column) 
for $N_f = 2,3,4,5$ and $6$, from top to bottom, respectively. 
Equations (\ref{chiralcontformula}) and (\ref{chiralcontformularatio}) 
are used and the resulting $\chi^2/dof$ is shown at the top of each plot. The various
colors correspond to different lattice spacings (i.e. different bare $\beta$) and were obtained
by interpolations of $a^2/w_0^2$ as a linear function of $m_\pi^2 w_0^2$ together with equations
(\ref{chiralcontformula}) and (\ref{chiralcontformularatio}); see text for more details.
}
\label{chiralcontfpimrhofigure}
\end{figure}

In the first one, at each $N_f$ the decay constant and the $\varrho$ mass are extrapolated separately to the
chiral-continuum limit in $w_0$ units. Concretely, the extrapolation is via
\bea
X w_0 = C_0 + C_1 m_\pi^2 w_0^2 + C_2 \frac{a^2}{w_0^2} + C_3 \frac{a^2}{w_0^2} m_\pi^2 w_0^2
\label{chiralcontformula}
\eea
where $X$ is either $m_\varrho$ or $f_\pi$ and there are four fit parameters $C_{0,1,2,3}$ 
hence $dof=12$ for each $N_f$. In this procedure we obtain $f_\pi w_0$ and
$m_\varrho w_0$ in the chiral-continuum limit at each $N_f$ and the results are given in table \ref{chiralcontfpimrho}
together with the $\chi^2/dof$ values. The extrapolations are shown in figure \ref{chiralcontfpimrhofigure} where
the chiral-continuum final results are shown in black together with the measured data. The solid lines corresponding
to each bare $\beta$ were obtained by fitting the scale $a^2/w_0^2$ as a linear function of $m_\pi^2 w_0^2$ together
with equation (\ref{chiralcontformula}). Clearly cut-off effects are small which is due to 
our choice of discretization and the choice of $w_0$ to set the scale.

In the second procedure the ratio
is fitted directly via
\bea
\frac{m_\varrho}{f_\pi} = C_0 + C_1 m_\pi^2 w_0^2 + C_2 \frac{a^2}{w_0^2}
\label{chiralcontformularatio}
\eea
where now we have three fit parameters $C_{0,1,2}$ hence $dof=13$ for all $N_f$. 
The results are shown again in figure \ref{chiralcontfpimrhofigure}. Clearly, both
cut-off and mass effects are remarkably small for the ratio although the mass-dependence of both $m_\varrho w_0$ and
$f_\pi w_0$ are much larger in comparison.

\begin{figure}
\begin{center}
\includegraphics[width=8cm]{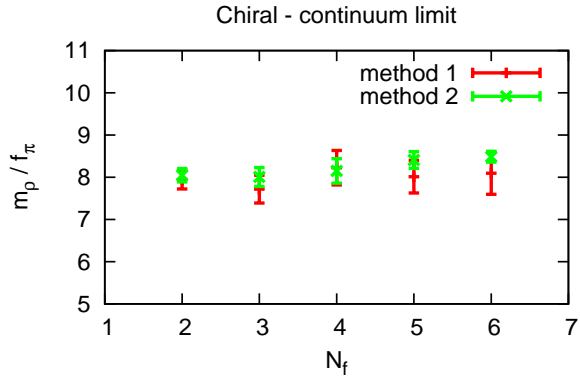}
\end{center}
\caption{Our final results for the ratio $m_\varrho / f_\pi$ in the chiral-continuum limit for each $N_f$. Two
procedures are used for the chiral-continuum extrapolation in order to assess systematic errors; the two agrees within
errors. Method one is our final result and method two serves as a cross-check or confirmation.}
\label{mrhofpinf}
\end{figure}

\begin{table}
\begin{center}
\begin{tabular}{|c|c|c|c|}
\hline
$N_f$ & $f_\pi w_0$ & $m_\varrho w_0$ & $m_\varrho / f_\pi$ \\
\hline
\hline
2 & 0.0801(5) & 0.64(2) & 7.9(2) \\
\hline
3 & 0.082(1) & 0.63(3) & 7.7(3) \\
\hline
4 & 0.0824(9) & 0.68(3) & 8.2(4) \\
\hline
5 & 0.086(1) & 0.69(3) & 8.0(4) \\
\hline
6 & 0.088(2) & 0.71(4) & 8.1(5) \\
\hline
\end{tabular}
\end{center}
\caption{
\label{chiralcontfpimrho}
Continuum results for each $N_f$ in the chiral limit.}
\end{table}

The above two procedures give compatible results for $m_\varrho / f_\pi$ 
within errors, the final results are shown in figure \ref{mrhofpinf}. The final errors are dominated by the errors on
$m_\varrho$, the errors on $f_\pi$ are negligible in comparison. We take the results from the first procedure as our 
final continuum results and use the second procedure, with its much smaller error, as confirmation or cross-check.

Our main conclusion can be drawn from figure \ref{mrhofpinf}; 
the $N_f$-dependence of the ratio $m_\varrho / f_\pi$ is remarkably small. We may even fit the results to a
constant and obtain acceptable statistical fits. Using the first procedure one obtains $m_\varrho / f_\pi = 7.95(15)$
whereas the second procedure leads to $m_\varrho / f_\pi = 8.28(8)$, with $\chi^2/dof = 0.26$ and $1.69$,
respectively. The agreement is within $2\,\sigma$.

This largely $N_f$-independent behavior is consistent with the
observation that at each $N_f$ the mass-dependence of the ratio is also very small. In addition note that free fermions
have $m_\varrho = 2m$ as is the case for any meson and also $f_\pi = \sqrt{12} m$ \cite{Cichy:2008gk}, leading to a very small
ratio $m_\varrho / f_\pi = 1 / \sqrt{3}$. This result in the free theory can be thought of as the relevant 
ratio at $N_f = 33/2$ at the upper end of the conformal window. Hence we conclude that well below the conformal
window, $2 \leq N_f \leq 6$ the ratio is relatively stable at around $\sim 8.0$ and somewhere in the range $7 \leq N_f
\leq 16.5$ it drops an order of magnitude to $\sim 0.6$. 
The extension of our work to this remaining range $7 \leq N_f \leq 16.5$
is left for future work and we believe that once it is completed it may serve as valuable
insight into the appearance of the conformal window, presumably at around $N_f \sim 13$. Note that inside the conformal
window the chiral limit of the ratio $m_\varrho / f_\pi$ is understood similarly to the free theory;
both the numerator and the denominator is finite at finite fermion mass with a well-defined ratio in the chiral limit.

\section{Conclusion and outlook}
\label{conclusionandoutlook}

In this work we have determined the $m_\varrho / f_\pi$ ratio in the chiral-continuum limit of $SU(3)$ gauge theory
coupled to $N_f = 2,3,4,5,6$ fermions in the fundamental representation in such a way that all systematic errors are
fully controlled. A remarkable $N_f$-independence is observed with $m_\varrho / f_\pi = 7.95(15)$ while several
quantities do show non-trivial $N_f$-dependence. The motivation for our study was that in a large class of strongly interacting
extensions of the Standard Model the experimentally most easily accessible new composite particle is the vector
resonance. The scale in these models is set by $f_\pi = 246\;GeV$ hence we are led to conclude that as long as the
gauge group is chosen to be $SU(3)$ the first resonance ought to be at around $\sim 2\;TeV$ independent of the specific fermion
content of the theory. 

Our result is not only a robust prediction for the mass of the vector resonance but also a host of other related
quantities. The KSRF-relations \cite{Kawarabayashi:1966kd, Riazuddin:1966sw} establish relationships among the vector
mass, its width $\Gamma_\varrho$ and decay constant $f_\varrho$ and the $\varrho\pi\pi$ coupling
$g_{\rho\pi\pi}$. Specifically,
\bea
g_{\varrho\pi\pi} &=& \frac{m_\varrho}{f_\varrho} = 
\sqrt{48\pi \frac{\Gamma_\varrho}{m_\varrho} } = \frac{1}{\sqrt{2}} \frac{m_\varrho}{f_\pi}\;.
\label{ksrfrelation}
\eea
The assumptions underlying the KSRF-relations are the applicability of leading order chiral perturbation theory,
vector meson dominance and vector meson universality\footnote{ 
It is worthwhile to point out that even though in QCD the KSRF-relations are only approximate at best
in supersymmetric QCD they have actually been rigorously derived \cite{Komargodski:2010mc}.
}. The last two conditions completely determine the way the vector
resonance ought to appear in the chiral Lagrangian and simple leading order calculations lead to the above relations;
see \cite{Bando:1987br} for a review. These relations are surprisingly accurate in QCD and it is expected that they
become even more accurate closer to the chiral limit. Hence our result for an approximately $N_f$-independent ratio
$m_\varrho / f_\pi$ leads to similar results for the coupling $g_{\varrho\pi\pi}$ and the $\varrho$ width and decay
constant in $m_\varrho$ units. In a strong dynamics inspired composite Higgs scenario this 
means $\Gamma_\varrho \sim 410\;GeV$, $f_\varrho \sim
348\;GeV$ and $g_{\varrho\pi\pi} \sim 5.62$, independently of the details of the fermion content as long as $SU(3)$ is
the gauge group. These additional results are especially useful because a direct lattice calculation of
$g_{\varrho\pi\pi}$ or $\Gamma_\varrho$ is very challenging.

Note that the KSRF-relations merely relate $m_\varrho / f_\pi$ to other quantities in the given theory
at fixed $N_f$. Hence even if the validity of the KSRF-relations is accepted it is not at all clear why this particular ratio
is insensitive to $N_f$ and it would be welcome to derive it at least approximately from first principles. 

The above is especially true since the change in
the detailed dynamics of gauge theory as $N_f$ is varied is highly non-trivial. As $N_f$ grows, the number of massless
particles increases, the running of the renormalized
coupling slows down, the $S$-parameter increases but $S/N_f$ decreases \cite{Appelquist:2010xv, Appelquist:2014zsa}, 
the topological susceptibility decreases in $m_\pi^2f_\pi^2$ units, the mass of
the $O^{++}$ scalar in $f_\pi$ units decreases \cite{Aoki:2013zsa, Aoki:2014oha, Rinaldi:2015axa, Fodor:2016pls,
Aoki:2016wnc, Appelquist:2016viq}, yet the vector meson related 
quantities stay roughly constant. It would be interesting to see how $m_\varrho / f_\pi$ changes across the lower end of the
conformal window, presumably close to $N_f \sim 13$ and how the free value $1/\sqrt{3}$ is reached at $N_f = 33/2$ at
the upper end of the conformal window. For this investigation the starting point must be the extension of the result
(\ref{mpil}) to $N_f \geq 7$ because it is not at all guaranteed that (\ref{mpil}) holds for flavor numbers beyond the range
considered in this work.

It should be noted that lattice results indicate that $m_\varrho / f_\pi$ is not completely universal, it does depend on
the gauge group. Evidence comes from $SU(2)$ simulations with $N_f = 2, 4$
fundamental fermions. With $N_f = 2$ continuum results \cite{Lewis:2011zb, Arthur:2016dir, Drach:2017btk} 
are available in the chiral limit, $m_\varrho / f_\pi \sim 15$
while with $N_f = 4$ results at finite lattice spacing \cite{Amato:2018nvj} 
indicate $m_\varrho / f_\pi > 10$. It would be worthwhile to
obtain fully controlled continuum results with $SU(2)$ at $N_f = 2,3,4$ and perhaps 
$N_f = 5$ in order to see whether the $N_f$-independence
below the conformal window we have seen for $SU(3)$ is also present with $SU(2)$ or not. In any case
a larger $m_\varrho / f_\pi$ ratio for $SU(2)$ relative to $SU(3)$ is in line with large-$N$ scaling arguments since
$m_\varrho \sim O(1)$ while $f_\pi \sim O(\sqrt{N})$; see \cite{Bali:2013kia, DeGrand:2016pur} and references therein.

\section*{Acknowledgements}

DN would like to thank Zoltan Fodor, Sandor Katz and Steve Sharpe for insightful discussions.
This work was supported by the NKFIH under the grant 
KKP-126769. Simulations were performed on the GPU clusters at Eotvos
University in Budapest, Hungary. We would like to thank Kalman Szabo for
code development.

\newpage

\section{Data tables}
\label{alldata}

\begin{table}[b]
\footnotesize
\begin{center}
\begin{tabular}{|c|c|c|c|c|c|}
\hline
$N_f$ & $\beta$ & $m$ & $L/a$ & $a m_\pi$ & $a f_\pi$ \\
\hline
\hline
2 & 3.92 & 0.0075 & 16 & 0.1650(8) & 0.0565(5) \\
\hline
 &  & & 20 & 0.1589(4) & 0.0602(3) \\
\hline
 &  & & 24 & 0.1577(4) & 0.0611(2) \\
\hline
 &  & & 28 & 0.1563(2) & 0.0613(2) \\
\hline
 &  & & $\infty$ & 0.1560(3) $\;\;$ 1.11 & 0.0616(2) $\;\;$ 0.48 \\
\hline
\hline
3 & 3.77 & 0.0110 & 16 & 0.199(1) & 0.0577(4) \\
\hline
 &  & & 20 & 0.1929(7) & 0.0603(4) \\
\hline
 &  & & 24 & 0.1916(8) & 0.0612(2) \\
\hline
 &  & & 28 & 0.1904(4) & 0.0612(2) \\
\hline
 &  & & $\infty$ & 0.1902(4) $\;\;$ 0.28 & 0.0613(2) $\;\;$ 0.24 \\
\hline
\hline
4 & 3.61 & 0.0088 & 16 & 0.190(1) & 0.0482(6) \\
\hline
 &  & & 20 & 0.1779(6) & 0.0535(2) \\
\hline
 &  & & 24 & 0.1748(5) & 0.0558(2) \\
\hline
 &  & & 28 & 0.1743(3) & 0.0559(1) \\
\hline
 &  & & $\infty$ & 0.1734(3) $\;\;$ 0.97 & 0.0563(2) $\;\;$ 2.62 \\
\hline
\hline
5 & 3.40 & 0.0085 & 16 & 0.202(2) & 0.0460(8) \\
\hline
 &  & & 20 & 0.185(1) & 0.0518(3) \\
\hline
 &  & & 24 & 0.1807(9) & 0.0543(3) \\
\hline
 &  & & 28 & 0.1797(3) & 0.0543(2) \\
\hline
 &  & & $\infty$ & 0.1788(6) $\;\;$ 0.15 & 0.0548(2) $\;\;$ 2.04 \\
\hline
\hline
6 & 3.31 & 0.0080 & 20 & 0.192(5) & 0.030(1) \\
\hline
 &  & & 24 & 0.163(2) & 0.0372(5) \\
\hline
 &  & & 28 & 0.1591(5) & 0.0392(3) \\
\hline
 &  & & 32 & 0.1577(8) & 0.0396(2) \\
\hline
 &  & & $\infty$ & 0.155(1) $\;\;$ 2.87 & 0.0403(3) $\;\;$ 0.87 \\
\hline
\hline
\end{tabular}
\end{center}
\caption{
Volume dependence of $m_\pi$ and $f_\pi$. The infinite volume extrapolated result is also shown together with
the $\chi^2/dof$ of the extrapolations; $dof = 2$.
\label{finvoltable}
}
\end{table}

\begin{table}
\footnotesize
\begin{center}
\begin{tabular}{|c|c|c|c|c|c|c|c|c|}
\hline
$N_f$ & $\beta$ & $m$ & $L/a$ & $am_\pi$ & $af_\pi$ & $a m_\varrho$ & $w_0/a$ & $10^4 a^4 \chi$ \\
\hline
\hline
\hline
2 & 3.84 & 0.0130 & 24 & 0.2221(1) & 0.1066(3) & 0.61(2)  &  1.140(1)  & 1.5(3) \\
\hline                                                                             
  &      & 0.0100 & 24 & 0.1957(2) & 0.1037(2) & 0.58(2)  &  1.144(1)  & 1.3(2) \\
\hline                                                                             
  &      & 0.0088 & 24 & 0.1839(2) & 0.1020(2) & 0.59(2)  &  1.1493(6) & 1.5(2) \\
\hline                                                                             
  &      & 0.0075 & 24 & 0.1704(2) & 0.1006(2) & 0.61(1)  &  1.151(1)  & 1.4(2) \\
\hline                                                                             
\hline                                                                             
  & 3.92 & 0.0147 & 20 & 0.2177(6) & 0.0929(3) & 0.548(6) &  1.337(3)  & 0.7(1) \\
\hline                                                                             
  &      & 0.0100 & 24 & 0.1812(3) & 0.0889(2) & 0.524(8) &  1.351(2)  & 0.47(6) \\
\hline                                                                             
  &      & 0.0088 & 24 & 0.1701(2) & 0.0885(2) & 0.517(6) &  1.350(2)  & 0.53(7) \\
\hline                                                                             
  &      & 0.0075 & 28 & 0.1563(2) & 0.0867(2) & 0.501(4) &  1.360(1)  & 0.60(6) \\
\hline                                                                             
\hline                                                                             
  & 4.03 & 0.0100 & 28 & 0.1624(4) & 0.0737(3) & 0.424(4) &  1.683(2)  & 0.24(4) \\
\hline                                                                             
  &      & 0.0088 & 28 & 0.1533(4) & 0.0720(4) & 0.420(6) &  1.689(3)  & 0.16(3) \\
\hline                                                                             
  &      & 0.0062 & 32 & 0.1300(9) & 0.0700(1) & 0.403(3) &  1.700(2)  & 0.21(4) \\
\hline                                                                             
  &      & 0.0050 & 32 & 0.1153(3) & 0.0686(2) & 0.402(6) &  1.708(2)  & 0.21(3) \\
\hline                                                                             
\hline                                                                             
  & 4.26 & 0.0115 & 28 & 0.1465(8) & 0.0526(3) & 0.314(3) &  2.50(1)   & 0.04(1) \\
\hline                                                                             
  &      & 0.0100 & 32 & 0.1367(5) & 0.0519(4) & 0.302(4) &  2.508(7)  & 0.043(9) \\
\hline                                                                             
  &      & 0.0088 & 32 & 0.1279(4) & 0.0500(2) & 0.302(2) &  2.550(7)  & 0.034(5) \\
\hline                                                                             
  &      & 0.0075 & 36 & 0.1173(4) & 0.0494(4) & 0.292(2) &  2.55(1)   & 0.05(1) \\
\hline                                                                                      
\hline                                                                                      
\hline                                                                                      
3 & 3.69 & 0.0158 & 20 & 0.2509(5) & 0.1061(3) & 0.614(6) &  1.175(2)  & 1.2(2) \\
\hline                                                                             
  &      & 0.0130 & 20 & 0.2271(4) & 0.1032(2) & 0.60(1)  &  1.189(2)  & 1.4(2) \\
\hline                                                                             
  &      & 0.0105 & 24 & 0.2036(3) & 0.1002(3) & 0.58(2)  &  1.200(1)  & 1.1(1) \\
\hline                                                                             
  &      & 0.0085 & 24 & 0.1849(4) & 0.0974(2) & 0.57(1)  &  1.206(1)  & 1.0(1) \\
\hline                                                                             
\hline                                                                             
  & 3.77 & 0.0140 & 24 & 0.2168(3) & 0.0898(2) & 0.53(1)  &  1.402(2)  & 0.65(6) \\
\hline                                                                             
  &      & 0.0110 & 24 & 0.1916(8) & 0.0865(3) & 0.47(1)  &  1.413(2)  & 0.54(8) \\
\hline                                                                             
  &      & 0.0095 & 24 & 0.1786(5) & 0.0840(3) & 0.49(2)  &  1.434(2)  & 0.55(7) \\
\hline                                                                             
  &      & 0.0075 & 28 & 0.1584(2) & 0.0828(2) & 0.48(1)  &  1.436(2)  & 0.35(6) \\
\hline                                                                             
\hline                                                                             
  & 3.86 & 0.0145 & 24 & 0.2012(6) & 0.0762(4) & 0.44(1)  &  1.694(5)  & 0.22(4) \\
\hline                                                                             
  &      & 0.0130 & 24 & 0.1906(4) & 0.0747(3) & 0.44(1)  &  1.698(6)  & 0.23(3) \\
\hline                                                                             
  &      & 0.0110 & 24 & 0.1750(7) & 0.0723(4) & 0.426(8) &  1.722(5)  & 0.18(3) \\
\hline                                                                             
  &      & 0.0095 & 28 & 0.1620(2) & 0.0715(2) & 0.418(5) &  1.732(4)  & 0.22(5) \\
\hline                                                                             
\hline                                                                             
  & 4.04 & 0.0150 & 24 & 0.177(1) & 0.0562(6) & 0.347(3)  &  2.38(2)   & 0.07(2) \\
\hline                                                                             
  &      & 0.0111 & 28 & 0.1509(6) & 0.0537(4) & 0.330(9) &  2.44(1)   & 0.07(1) \\
\hline                                                                             
  &      & 0.0085 & 32 & 0.1314(7) & 0.0505(3) & 0.304(8) &  2.50(1)   & 0.05(1) \\
\hline                                                                             
  &      & 0.0067 & 36 & 0.1168(5) & 0.0497(3) & 0.283(8) &  2.498(8)  & 0.022(5) \\
\hline                                                                                      
\hline                                                                                      
\hline                                                                                      
4 & 3.54 & 0.0140 & 20 & 0.2378(5) & 0.0981(3) & 0.566(9) &  1.304(3)  & 0.8(1) \\
\hline                                                                             
  &      & 0.0120 & 24 & 0.2198(4) & 0.0952(2) & 0.56(2)  &  1.323(2)  & 0.8(1) \\
\hline                                                                             
  &      & 0.0088 & 28 & 0.1882(3) & 0.0903(2) & 0.51(1)  &  1.344(2)  & 0.8(1) \\
\hline                                                                             
  &      & 0.0062 & 32 & 0.1589(2) & 0.0849(2) & 0.48(1)  &  1.375(2)  & 0.6(1) \\
\hline                                                                             
\hline                                                                             
  & 3.61 & 0.0146 & 20 & 0.2269(7) & 0.0860(3) & 0.47(2)  &  1.511(5)  & 0.51(7) \\
\hline                                                                             
  &      & 0.0110 & 24 & 0.1955(5) & 0.0814(3) & 0.46(1)  &  1.546(5)  & 0.28(5) \\
\hline                                                                             
  &      & 0.0088 & 28 & 0.1743(3) & 0.0790(2) & 0.44(1)  &  1.570(2)  & 0.41(3) \\
\hline                                                                             
  &      & 0.0075 & 32 & 0.1609(1) & 0.0767(1) & 0.43(1)  &  1.583(2)  & 0.34(4) \\
\hline                                                                             
\hline                                                                             
  & 3.71 & 0.0151 & 24 & 0.2062(7) & 0.0727(4) & 0.411(9) &  1.849(9)  & 0.22(3) \\
\hline                                                                             
  &      & 0.0121 & 28 & 0.1846(4) & 0.0693(3) & 0.398(6) &  1.888(7)  & 0.16(4) \\
\hline                                                                             
  &      & 0.0088 & 32 & 0.1572(3) & 0.0654(2) & 0.388(7) &  1.936(5)  & 0.13(2) \\
\hline                                                                             
  &      & 0.0080 & 32 & 0.1490(3) & 0.0639(2) & 0.37(2)  &  1.972(6)  & 0.12(2) \\
\hline                                                                             
\hline                                                                             
  & 3.88 & 0.0150 & 28 & 0.1774(7) & 0.0546(5) & 0.326(5) &  2.58(2)   & 0.06(1) \\
\hline                                                                             
  &      & 0.0130 & 28 & 0.163(1)  & 0.0520(5) & 0.303(5) &  2.65(2)   & 0.032(6) \\
\hline                                                                             
  &      & 0.0110 & 32 & 0.1501(6) & 0.0505(3) & 0.298(4) &  2.71(1)   & 0.022(3) \\
\hline                                                                             
  &      & 0.0088 & 36 & 0.1324(8) & 0.0483(2) & 0.284(4) &  2.74(1)   & 0.030(4) \\
\hline
\hline
\end{tabular}
\end{center}
\caption{
\label{data234}
Data for $N_f = 2,3,4$.}
\end{table}

\begin{table}
\footnotesize
\begin{center}
\begin{tabular}{|c|c|c|c|c|c|c|c|c|}
\hline
$N_f$ & $\beta$ & $m$ & $L/a$ & $am_\pi$ & $af_\pi$ & $a m_\varrho$ & $w_0/a$ & $10^4 a^4 \chi$ \\
\hline
\hline
\hline
5 & 3.33 & 0.0190 & 20 & 0.2888(4) & 0.1070(4) & 0.60(1)   & 1.314(3)   &  1.3(3) \\
\hline                                                                               
  &      & 0.0148 & 20 & 0.2553(4) & 0.0995(4) & 0.57(2)   & 1.367(5)   &  0.9(1) \\
\hline                                                                               
  &      & 0.0105 & 24 & 0.2144(3) & 0.0917(2) & 0.53(1)   & 1.416(3)   &  0.8(1) \\
\hline                                                                               
  &      & 0.0081 & 28 & 0.1894(2) & 0.0861(2) & 0.498(6)  & 1.456(2)   &   0.67(7) \\
\hline                                                                               
\hline                                                                               
  & 3.40 & 0.0168 & 20 & 0.2555(5) & 0.0915(4) & 0.522(5)  & 1.527(4)   &   0.7(1) \\
\hline                                                                               
  &      & 0.0131 & 24 & 0.2242(5) & 0.0855(2) & 0.486(2)  & 1.594(4)   &   0.42(7) \\
\hline                                                                               
  &      & 0.0093 & 28 & 0.1884(5) & 0.0783(4) & 0.449(4)  & 1.665(4)   &   0.43(6) \\
\hline                                                                               
  &      & 0.0075 & 32 & 0.1690(3) & 0.0746(2) & 0.430(5)  & 1.704(3)   &   0.39(9) \\
\hline                                                                                        
\hline                                                                                        
  & 3.51 & 0.0174 & 24 & 0.2337(8) & 0.0771(5) & 0.443(4)  & 1.88(1)    &   0.18(2) \\
\hline                                                                                
  &      & 0.0142 & 24 & 0.2104(7) & 0.0713(3) & 0.419(5)  & 1.966(9)   &   0.21(4) \\
\hline                                                                                
  &      & 0.0110 & 28 & 0.1845(4) & 0.0662(4) & 0.391(5)  & 2.053(7)   &   0.14(2) \\
\hline                                                                                
  &      & 0.0079 & 32 & 0.1545(3) & 0.0624(2) & 0.360(9)  & 2.117(7)   &   0.13(3) \\
\hline                                                                                
\hline                                                                                
  & 3.68 & 0.0153 & 28 & 0.1874(6) & 0.0562(3) & 0.329(5)  & 2.65(1)    &  0.06(2)  \\
\hline                                                                                
  &      & 0.0135 & 28 & 0.1770(9) & 0.0523(4) & 0.313(6)  & 2.76(2)    &  0.028(4) \\
\hline                                                                                
  &      & 0.0104 & 32 & 0.155(1) & 0.0500(3) & 0.294(3)   & 2.85(2)    &  0.034(6) \\
\hline                                                                                
  &      & 0.0082 & 36 & 0.1347(8) & 0.0469(3) & 0.268(5)  & 2.95(2)    &  0.023(4) \\
\hline                                                                                        
\hline                                                                                        
\hline                                                                                        
6 & 3.12 & 0.0192 & 20 & 0.2947(5) & 0.1025(3) & 0.569(3)  & 1.471(5)   &   0.6(1) \\
\hline                                                                               
  &      & 0.0150 & 24 & 0.2590(3) & 0.0945(2) & 0.534(5)  & 1.547(4)   &   0.9(1) \\
\hline                                                                               
  &      & 0.0117 & 24 & 0.2283(4) & 0.0866(4) & 0.497(6)  & 1.628(6)   &   0.6(1) \\
\hline                                                                               
  &      & 0.0086 & 28 & 0.1960(3) & 0.07939(7) & 0.458(5) & 1.710(5)   &   0.51(8) \\
\hline                                                                               
\hline                                                                               
  & 3.19 & 0.0150 & 24 & 0.2442(6) & 0.0842(3) & 0.476(3)  & 1.763(6)   &   0.42(5) \\
\hline                                                                               
  &      & 0.0120 & 28 & 0.2171(3) & 0.0783(3) & 0.441(4)  & 1.861(7)   &   0.30(6) \\
\hline                                                                               
  &      & 0.0100 & 28 & 0.1981(4) & 0.0739(3) & 0.420(5)  & 1.924(6)   &   0.29(4) \\
\hline                                                                               
  &      & 0.0085 & 32 & 0.1824(3) & 0.0706(1) & 0.405(3)  & 1.969(5)   &   0.31(5) \\
\hline                                                                               
\hline                                                                               
  & 3.31 & 0.0150 & 28 & 0.2187(8) & 0.0701(2) & 0.398(2)  & 2.190(8)   &   0.19(3) \\
\hline                                                                               
  &      & 0.0125 & 28 & 0.203(1) & 0.0667(5) & 0.382(3)   & 2.28(1)    &   0.12(2) \\
\hline                                                                               
  &      & 0.0095 & 32 & 0.1731(2) & 0.0602(2) & 0.344(3)  & 2.42(1)    &   0.14(2) \\
\hline                                                                               
  &      & 0.0085 & 36 & 0.1636(3) & 0.0584(1) & 0.34(1)   & 2.452(7)   &   0.11(2) \\
\hline                                                                               
\hline                                                                               
  & 3.41 & 0.0130 & 32 & 0.1860(5) & 0.0581(2) & 0.328(2)  & 2.663(9)   &   0.056(7) \\
\hline                                                                               
  &      & 0.0112 & 32 & 0.1719(6) & 0.0542(3) & 0.319(5)  & 2.80(1)    &   0.042(5) \\
\hline                                                                               
  &      & 0.0100 & 36 & 0.1618(3) & 0.0525(2) & 0.292(3)  & 2.880(9)   &   0.05(1) \\
\hline                                                                               
  &      & 0.0089 & 36 & 0.1513(5) & 0.0507(3) & 0.284(3)  & 2.92(1)    &   0.045(9) \\
\hline
\hline
\end{tabular}
\end{center}
\caption{
\label{data56}
Data for $N_f = 5,6$.}
\end{table}


\begin{thebibliography}{99}

\footnotesize

\bibitem{Fodor:2009wk} 
  Z.~Fodor, K.~Holland, J.~Kuti, D.~Nogradi and C.~Schroeder,
  Phys.\ Lett.\ B {\bf 681}, 353 (2009)
  [arXiv:0907.4562 [hep-lat]].



\bibitem{Jin:2009mc} 
  X.~Y.~Jin and R.~D.~Mawhinney,
  PoS LAT {\bf 2009}, 049 (2009)
  [arXiv:0910.3216 [hep-lat]].



\bibitem{Aoki:2013xza} 
  Y.~Aoki {\it et al.} [LatKMI Collaboration],
  Phys.\ Rev.\ D {\bf 87}, no. 9, 094511 (2013)
  [arXiv:1302.6859 [hep-lat]].



\bibitem{Jin:2013hpa} 
  X.~Y.~Jin and R.~D.~Mawhinney,
  arXiv:1304.0312 [hep-lat].



\bibitem{Fleming:2013tra} 
  G.~T.~Fleming {\it et al.} [LSD Collaboration],
  arXiv:1312.5298 [hep-lat].



\bibitem{Appelquist:2016viq} 
  T.~Appelquist {\it et al.},
  Phys.\ Rev.\ D {\bf 93}, no. 11, 114514 (2016)
  [arXiv:1601.04027 [hep-lat]].



\bibitem{Appelquist:2018yqe} 
  T.~Appelquist {\it et al.} [Lattice Strong Dynamics Collaboration],
  Phys.\ Rev.\ D {\bf 99}, no. 1, 014509 (2019)
  [arXiv:1807.08411 [hep-lat]].



\bibitem{Fodor:2012ty} 
  Z.~Fodor, K.~Holland, J.~Kuti, D.~Nogradi, C.~Schroeder and C.~H.~Wong,
  Phys.\ Lett.\ B {\bf 718}, 657 (2012)
  [arXiv:1209.0391 [hep-lat]].



\bibitem{Fodor:2016pls} 
  Z.~Fodor, K.~Holland, J.~Kuti, S.~Mondal, D.~Nogradi and C.~H.~Wong,
  PoS LATTICE {\bf 2015}, 219 (2016)
  [arXiv:1605.08750 [hep-lat]].



\bibitem{Fodor:2016wal} 
  Z.~Fodor, K.~Holland, J.~Kuti, S.~Mondal, D.~Nogradi and C.~H.~Wong,
  Phys.\ Rev.\ D {\bf 94}, no. 1, 014503 (2016)
  [arXiv:1601.03302 [hep-lat]].



\bibitem{Kawarabayashi:1966kd} 
  K.~Kawarabayashi and M.~Suzuki,
  Phys.\ Rev.\ Lett.\  {\bf 16}, 255 (1966).



\bibitem{Riazuddin:1966sw} 
  Riazuddin and Fayyazuddin,
  Phys.\ Rev.\  {\bf 147}, 1071 (1966).



\bibitem{Cichy:2008gk} 
  K.~Cichy, J.~Gonzalez Lopez, K.~Jansen, A.~Kujawa and A.~Shindler,
  Nucl.\ Phys.\ B {\bf 800}, 94 (2008)
  [arXiv:0802.3637 [hep-lat]].



\bibitem{Morningstar:2003gk} 
  C.~Morningstar and M.~J.~Peardon,
  Phys.\ Rev.\ D {\bf 69}, 054501 (2004)
  [hep-lat/0311018].



\bibitem{Durr:2010aw} 
  S.~Durr {\it et al.},
  JHEP {\bf 1108}, 148 (2011)
  [arXiv:1011.2711 [hep-lat]].



\bibitem{Fodor:2012td} 
  Z.~Fodor, K.~Holland, J.~Kuti, D.~Nogradi and C.~H.~Wong,
  JHEP {\bf 1211}, 007 (2012)
  [arXiv:1208.1051 [hep-lat]].



\bibitem{Fodor:2015baa} 
  Z.~Fodor, K.~Holland, J.~Kuti, S.~Mondal, D.~Nogradi and C.~H.~Wong,
  JHEP {\bf 1506}, 019 (2015)
  [arXiv:1503.01132 [hep-lat]].



\bibitem{Fodor:2015zna} 
  Z.~Fodor, K.~Holland, J.~Kuti, S.~Mondal, D.~Nogradi and C.~H.~Wong,
  JHEP {\bf 1509}, 039 (2015)
  [arXiv:1506.06599 [hep-lat]].



\bibitem{Fodor:2017die} 
  Z.~Fodor, K.~Holland, J.~Kuti, D.~Nogradi and C.~H.~Wong,
  EPJ Web Conf.\  {\bf 175}, 08027 (2018)
  [arXiv:1711.04833 [hep-lat]].



\bibitem{Duane:1987de} 
  S.~Duane, A.~D.~Kennedy, B.~J.~Pendleton and D.~Roweth,
  Phys.\ Lett.\ B {\bf 195}, 216 (1987).



\bibitem{Clark:2006fx} 
  M.~A.~Clark and A.~D.~Kennedy,
  Phys.\ Rev.\ Lett.\  {\bf 98}, 051601 (2007)
  [hep-lat/0608015].



\bibitem{Sexton:1992nu} 
  J.~C.~Sexton and D.~H.~Weingarten,
  Nucl.\ Phys.\ B {\bf 380}, 665 (1992).



\bibitem{Takaishi:2005tz} 
  T.~Takaishi and P.~de Forcrand,
  Phys.\ Rev.\ E {\bf 73}, 036706 (2006)
  [hep-lat/0505020].



\bibitem{Borsanyi:2012zs} 
  S.~Borsanyi {\it et al.},
  JHEP {\bf 1209}, 010 (2012)
  [arXiv:1203.4469 [hep-lat]].



\bibitem{Fodor:2014cpa} 
  Z.~Fodor, K.~Holland, J.~Kuti, S.~Mondal, D.~Nogradi and C.~H.~Wong,
  JHEP {\bf 1409}, 018 (2014)
  [arXiv:1406.0827 [hep-lat]].


\bibitem{Luscher:1990ux} 
  M.~Luscher,
  Nucl.\ Phys.\ B {\bf 354}, 531 (1991).


\bibitem{Luscher:1991cf} 
  M.~Luscher,
  Nucl.\ Phys.\ B {\bf 364}, 237 (1991).



\bibitem{Gasser:1986vb} 
  J.~Gasser and H.~Leutwyler,
  Phys.\ Lett.\ B {\bf 184}, 83 (1987).



\bibitem{Colangelo:2005gd} 
  G.~Colangelo, S.~Durr and C.~Haefeli,
  Nucl.\ Phys.\ B {\bf 721}, 136 (2005)
  [hep-lat/0503014].



\bibitem{Gasser:1984gg} 
  J.~Gasser and H.~Leutwyler,
  Nucl.\ Phys.\ B {\bf 250}, 465 (1985).



\bibitem{Leutwyler:1992yt} 
  H.~Leutwyler and A.~V.~Smilga,
  Phys.\ Rev.\ D {\bf 46}, 5607 (1992).



\bibitem{Luscher:2010iy} 
  M.~Lüscher,
  JHEP {\bf 1008}, 071 (2010)
  Erratum: [JHEP {\bf 1403}, 092 (2014)]
  [arXiv:1006.4518 [hep-lat]].



\bibitem{Billeter:2004wx} 
  B.~Billeter, C.~E.~Detar and J.~Osborn,
  Phys.\ Rev.\ D {\bf 70}, 077502 (2004)
  [hep-lat/0406032].



\bibitem{Bando:1987br} 
  M.~Bando, T.~Kugo and K.~Yamawaki,
  Phys.\ Rept.\  {\bf 164}, 217 (1988).


\bibitem{Komargodski:2010mc} 
  Z.~Komargodski,
  JHEP {\bf 1102}, 019 (2011)
  [arXiv:1010.4105 [hep-th]].


\bibitem{Appelquist:2010xv} 
  T.~Appelquist {\it et al.} [LSD Collaboration],
  Phys.\ Rev.\ Lett.\  {\bf 106}, 231601 (2011)
  [arXiv:1009.5967 [hep-ph]].



\bibitem{Appelquist:2014zsa} 
  T.~Appelquist {\it et al.} [LSD Collaboration],
  Phys.\ Rev.\ D {\bf 90}, no. 11, 114502 (2014)
  [arXiv:1405.4752 [hep-lat]].



\bibitem{Aoki:2013zsa} 
  Y.~Aoki {\it et al.} [LatKMI Collaboration],
  Phys.\ Rev.\ Lett.\  {\bf 111}, no. 16, 162001 (2013)
  [arXiv:1305.6006 [hep-lat]].



\bibitem{Aoki:2014oha} 
  Y.~Aoki {\it et al.} [LatKMI Collaboration],
  Phys.\ Rev.\ D {\bf 89}, 111502 (2014)
  [arXiv:1403.5000 [hep-lat]].



\bibitem{Rinaldi:2015axa} 
  E.~Rinaldi [LSD Collaboration],
  Int.\ J.\ Mod.\ Phys.\ A {\bf 32}, no. 35, 1747002 (2017)
  [arXiv:1510.06771 [hep-lat]].



\bibitem{Aoki:2016wnc} 
  Y.~Aoki {\it et al.} [LatKMI Collaboration],
  Phys.\ Rev.\ D {\bf 96}, no. 1, 014508 (2017)
  [arXiv:1610.07011 [hep-lat]].



\bibitem{Lewis:2011zb} 
  R.~Lewis, C.~Pica and F.~Sannino,
  Phys.\ Rev.\ D {\bf 85}, 014504 (2012)
  [arXiv:1109.3513 [hep-ph]].



\bibitem{Arthur:2016dir} 
  R.~Arthur, V.~Drach, M.~Hansen, A.~Hietanen, C.~Pica and F.~Sannino,
  Phys.\ Rev.\ D {\bf 94}, no. 9, 094507 (2016)
  [arXiv:1602.06559 [hep-lat]].



\bibitem{Drach:2017btk} 
  V.~Drach, T.~Janowski and C.~Pica,
  EPJ Web Conf.\  {\bf 175}, 08020 (2018)
  [arXiv:1710.07218 [hep-lat]].



\bibitem{Amato:2018nvj} 
  A.~Amato, V.~Leino, K.~Rummukainen, K.~Tuominen and S.~Tähtinen,
  arXiv:1806.07154 [hep-lat].


\bibitem{Bali:2013kia} 
  G.~S.~Bali, F.~Bursa, L.~Castagnini, S.~Collins, L.~Del Debbio, B.~Lucini and M.~Panero,
  JHEP {\bf 1306}, 071 (2013)
  [arXiv:1304.4437 [hep-lat]].


\bibitem{DeGrand:2016pur} 
  T.~DeGrand and Y.~Liu,
  Phys.\ Rev.\ D {\bf 94}, no. 3, 034506 (2016)
  Erratum: [Phys.\ Rev.\ D {\bf 95}, no. 1, 019902 (2017)]
  [arXiv:1606.01277 [hep-lat]].




\end{thebibliography}
\end{document}